\documentclass[pra]{revtex4}
\usepackage{textcomp}
\usepackage{mathtools, ulem, graphicx}
\usepackage[export]{adjustbox}
\usepackage{sidecap}
\usepackage{verbatim}
\usepackage{subfig}

\newcommand{\e}{\begin{equation*}\begin{aligned}}
\newcommand{\ee}{\end{aligned}\end{equation*}}

\newcommand{\en}{\begin{equation}\begin{aligned}}
\newcommand{\een}{\end{aligned} \end{equation}}

\newcommand{\f}[2]{\frac{#1}{#2}}

\newcommand{\ma}{\mathcal}

\newcommand{\pma}{\begin{pmatrix}}
\newcommand{\epma}{\end{pmatrix}}

\begin{document}
\scriptsize This is the Accepted Manuscript version of an article accepted for publication in European Journal of Physics. Neither the European Physical Society nor IOP Publishing Ltd is responsible for any errors or omissions in this version of the manuscript or any version derived from it. The Version of Record is available online at  https://doi.org/10.1088/1361-6404/ab895d.\\
\newline \normalsize
\title{Chirality Through Classical Physics}
\author{Chris L. Lin}
\affiliation{Department of Physics, University of Houston, Houston, TX 77204-5005}

\date{\today}


\begin{abstract}
Chirality, or handedness, is a topic that is common in biology and chemistry, yet is rarely discussed in physics courses. We provide a way of introducing the topic in classical physics, and demonstrate the merits of its inclusion -- such as a simple way to visually introduce the concept of symmetries in physical law -- along with giving some simple proofs using only basic matrix operations, thereby avoiding the full formalism of the three-dimensional point group.
\end{abstract}

\maketitle

\section{Introduction}

Chirality is a topic that spans several sciences, from the helicity of DNA in biology to the existence of organic enantiomers in chemistry \cite{chiralArc}, to the optical rotation of liquid crystal displays \cite{doi:10.1119/1.17801,opticalAc} and the handedness of radioactive decay in physics \cite{weak}, to name a few. It is a common phenomena of everyday life, from right-handed threaded screws and twisted ropes to the design choices made in the manufacture of objects catered to a majority right-handed population. As an illustration of the importance of the topic, it has appeared multiple times in the popular BBC-televised annual Christmas lectures given by the Royal Institution of Great Britain \cite{christmas1,christmas2}.\\

A natural starting point to discuss chirality is in optics when discussing the plane mirror, as chirality can broadly be defined as the study of the effects of replacing an object by its mirror image. The perennial question of a plane mirror's left-right inversion can be used to define chirality and introduce reflection and rotation matrices: this is done in section \ref{springboard}. Having introduced these matrices, in section \ref{math} we prove some simple facts about reflections and rotations using matrix multiplication, and introduce improper rotations. In section \ref{consequences}, we consider the effect of chirality on interactions, using knots as an example, and make a distinction between the behavior of chiral objects in physics versus whether the laws of physics are themselves chiral by discussing the violation of parity in the weak interaction. In section \ref{conclusions}, we summarize the merits of introducing chirality prior to a course in quantum mechanics, along with conclusions.

\section{Chirality Definition}\label{springboard}

\begin{center}
\begin{figure}[h]
\includegraphics[scale=0.5]{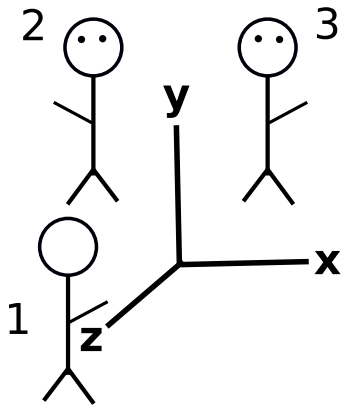}  \caption{The mirror image of the person is attained directly by the front-back inversion ($1 \rightarrow 3$), or by getting the image to face the person in the usual way with a rotation ($1 \rightarrow 2$), followed by noticing the result is off by a reflection in lateral direction ($2 \rightarrow 3$). The mirror sits in the xy-plane and the person is staring in the negative z-direction.}
\label{one}
\end{figure}
\end{center}

In geometric optics, it is commonly derived that the image is the same height as the object, upright, and at the same distance from the mirror as the object. Nevertheless it would be a mistake to say that the image is unchanged by the mirror. The question of the left-right inversion of a plane mirror, which has been discussed extensively in the literature \cite{ford3,Gee_1988,Kalmus_1989,3,4,ford2,ford4}, can be modeled with matrices, where $\ma R_i$ represents a reflection in the plane perpendicular to axis $i$, and $R_i(\theta)$ represents a rotation by angle $\theta$ about axis $i$:

\en \label{matrixmulteq}
\ma R_x R_y(\pi)&=\ma R_z\\
\pma
-1& 0 &0\\
0&1&0 \\
0 & 0 & 1
\epma
\pma
\cos \pi& 0 &\sin \pi\\
0&1&0 \\
-\sin \pi & 0 & \cos \pi
\epma
&=
\pma
1& 0 &0\\
0&1&0 \\
0 & 0 & -1
\epma
\een

which mathematically models the sequence of operations shown in Fig. \ref{one}.
The frontal inversion $z \rightarrow -z$ on the RHS is the simplest \textit{mathematical} way to get an object to face ourselves. However, the most common \textit{physical} way to get an object to face us is to rotate the object 180 degrees about the y-axis. We have a keen sense of symmetry and notice that the result after such a rotation differs from the mirror image by a reflection in the x-direction, hence the nomenclature that the mirror inverts left-right \footnote{For example, in chemistry, one configuration of a molecule may be designated as right-handed, and its mirror image is then left-handed.}. 
 We note that in Fig. \ref{one} the object can be distinguished from its mirror image, i.e., cannot be superposed on its mirror image after any rotation. Such an object is called chiral, a term first coined by Lord Kelvin \cite{kelvin1894molecular}. In chemistry the pair comprising the object and its mirror image are called enantiomers. \\

If an object cannot be distinguished from its mirror image, then it is called achiral. For example, if the person in Fig. \ref{one} rests his right hand at his side, then the object is achiral. In the next section we will derive the most general condition for an object to be achiral, but one can see that for this particular example, i.e. for the person with his hand at his side which we represent as a collection of points $X_{\text{person}}=\{(x,y,z)\}^T$, then due to the bilateral symmetry of the person,

\en
\pma
-1& 0 &0\\
0&1&0 \\
0 & 0 & 1
\epma 
\begin{Bmatrix}
\pma 
x\\y\\z
\epma
\end{Bmatrix}=
\begin{Bmatrix}
\pma 
x\\y\\z
\epma
\end{Bmatrix},
\een

i.e. the reflection does not change the set of points that comprise the person, and so $\ma R_x R_y(\pi)X_\text{person}=\ma R_zX_\text{person}$ becomes just  $R_y(\pi)X_\text{person}=\ma R_zX_\text{person}$, or that the mirror image ($\ma R_zX_\text{person}$) is just a rotation of the object ($R_y(\pi)X_\text{person}$), hence the object is achiral. Any object that has a plane of symmetry is achiral, but not all achiral objects have a plane of symmetry. In the next section we proceed to derive the most general condition for an object to be achiral: if an object is invariant under a rotation about an axis followed by a reflection in the plane perpendicular to the \textit{same} axis, a combined operation specified by a single axis called an improper rotation, then it is achiral.\\

\begin{center}
\begin{figure}[h] 
\includegraphics[scale=0.5]{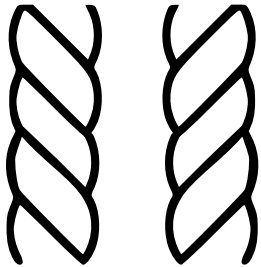}
\caption{Left- and right-handed helices, demonstrated with twisted rope. The left-handed helix on the left cannot be rotated into its mirror image, the right-handed helix on the right, hence is chiral.}
\label{ropeseg}
\end{figure}
\end{center}

The quintessential example of a chiral object that often appears in science is the helix. In undergraduate physics, examples include screws, twisted rope, solenoids, kinematics of a charge particles in magnetic fields, circular polarization, and the very definition of right-handed coordinate systems. A helix is right-handed if curling the fingers of your right-hand around the turns advances you in the direction of your thumb. Alternatively, borrowing from the rope and textile sector, a helix is right-handed if when laid vertically and facing you, the turns are moving from the bottom left to the upper right, which is called a `Z' twist due to the shape of the middle of that letter (see Fig. \ref{ropeseg}). A left-handed helix is called an `S' twist for similar reasons.

\section{Rotations and Reflections}\label{math}

Group theory is generally not part of the standard undergraduate curriculum. However, most students are familiar with rotation matrices. Such matrices allows one to be precise about the relation between the object and its mirror image instead of relying on descriptors such as left, right, turn, and indeed were already used for that purpose in the previous section, where the slipperiness of words are replaced with the clarity of maths. \\

The familiar rotation matrices are given by

\en \label{rotMatrices}
R_x(\alpha)=\pma1&0&0\\
0& \cos \alpha & -\sin \alpha \\
0&\sin \alpha & \cos \alpha
\epma,
R_y(\alpha)=\pma
\cos \alpha &0& \sin \alpha \\
0&1&0\\
-\sin \alpha &0& \cos \alpha
\epma,
R_z(\alpha)=\pma \cos \alpha & -\sin \alpha & 0\\
\sin \alpha & \cos \alpha & 0\\
0&0&1
\epma,
\een

each of which has determinant equal to one. We construct a general rotation of angle $\alpha$ about an axis in the $(\theta, \phi)$ direction, where $\theta$ is the polar angle and $\phi$ the azimuthal angle, using a similarity transformation:

\en \label{generalRot}
R_{\theta,\phi}(\alpha)=R_z(\phi)R_y(\theta)R_z(\alpha)R_y(-\theta) R_z(-\phi),
\een

which can be understood as rotating the $(\theta, \phi)$ axis (along with the point that's rotating about this axis) so that  $(\theta, \phi)$ aligns with the $z$-axis, then performing the rotation of $\alpha$ about the $z$-axis, and then undoing the initial rotation. The determinant of such a matrix is the product of determinants each of which has a value equal to $1$, so we have proven that any rotation has its determinant equal to $1$.\\

The reflection matrices are:

\en
\ma R_x=\pma -1&0&0\\
0&1&0\\
0&0&1
\epma,
\ma R_y=\pma
1&0&0\\
0&-1&0\\
0&0&1
\epma,
\ma R_z=\pma 1&0&0\\
0&1&0\\
0&0&-1
\epma,
\een

each of which has determinant equal to negative one compared to their counterparts in \eqref{rotMatrices}. The fact that reflections have a different determinant than rotations shows that these are in general different operations, although as we have seen, acting on achiral objects they give the same effect. \\

A general reflection about a plane whose normal is in the $(\theta, \phi)$ direction is given by an argument similar to \eqref{generalRot}:

\en
\ma R_{\theta,\phi}=R_z(\phi)R_y(\theta)\ma R_z R_y(-\theta) R_z(-\phi).
\een

The determinant of such a matrix is the product of determinants which is $-1$, so we have proven that any reflection has its determinant equal to $-1$.\\

A useful identity relates the parity operator $\ma P=\pma -1&0&0\\
0&-1&0\\
0&0&-1
\epma$ to reflections:

\en
\ma P R_{\theta,\phi}(180)=\ma R_{\theta,\phi}.
\een

We can now state the most general condition for an object $X=\begin{Bmatrix}
\pma x\\y\\z \epma \end{Bmatrix}$ to be achiral. If a reflection of the object is equivalent to a rotation

\en \label{condition}
\ma R_{\theta,\phi}X=R_{\theta',\phi'}(\alpha')X,
\een

then clearly $\ma R_{\theta,\phi}X$ can be superposed on $X$ via the subsequent rotation $R^{-1}_{\theta',\phi'}(\alpha')=R_{\theta',\phi'}(-\alpha')$, so that the object is achiral. However, the common way of defining achirality is if there exists an axis such that rotation about this axis, followed by a reflection in the plane perpendicular to this \textit{same} axis, leaves the object invariant. This is known as an improper- or roto-rotation. This can again be proven with just matrices. Starting with condition \eqref{condition} and using $\ma R_{\theta,\phi}^2=1$:

\en \label{rotorotation}
X&=\ma R_{\theta,\phi}R_{\theta',\phi'}(\alpha')X\\
&=\ma P R_{\theta,\phi} (180)R_{\theta',\phi'}(\alpha')X\\
&=\ma P R_{\theta'',\phi''}(\alpha'')X\\
&=\ma P R_{\theta'',\phi''} (180) R_{\theta'',\phi''}(180)  R_{\theta'',\phi''}(\alpha'')X\\
&=\ma R_{\theta'',\phi''}R_{\theta'',\phi''}(\alpha''+180)X\\
&=\ma R_{\theta'',\phi''}R_{\theta'',\phi''}(\beta)X.
\een

\begin{center}
\begin{figure}[h] 
\subfloat[A molecule that has no plane of symmetry, but can be rotated into its mirror image. It is invariant under parity, and the origin about which it is invariant is called the inversion center. Formally this is $S_2$ isometry. \label{paritymolecule}]
{\includegraphics[scale=0.22]{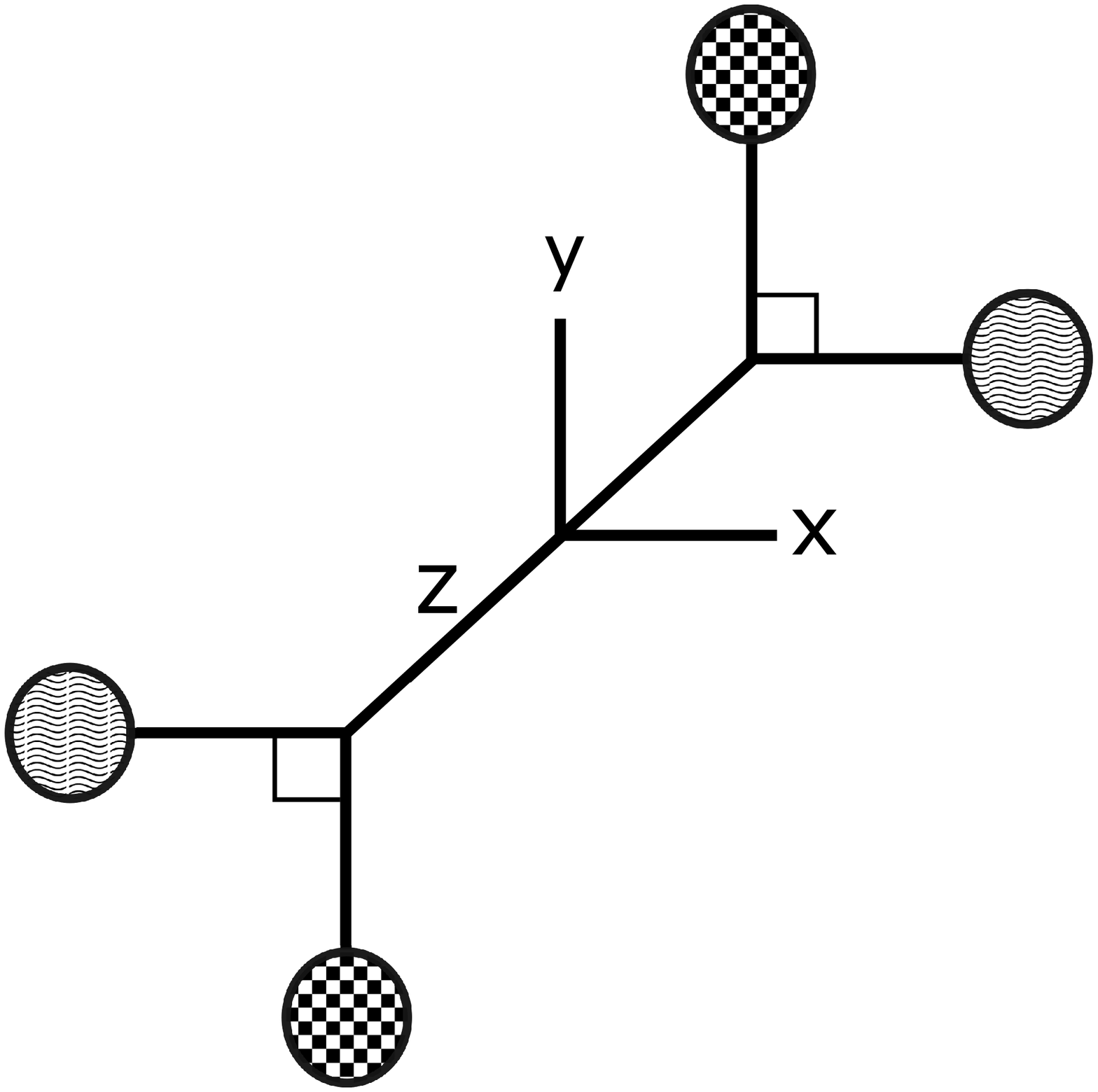}}  \hspace{60 pt}
\subfloat[A molecule that has no plane of symmetry or inversion center. A rotation of $90$ degrees (about the dotted line), followed by a reflection, leaves the molecule invariant. Formally this is $S_4$ isometry. \label{methane}]
{\includegraphics[scale=.3]{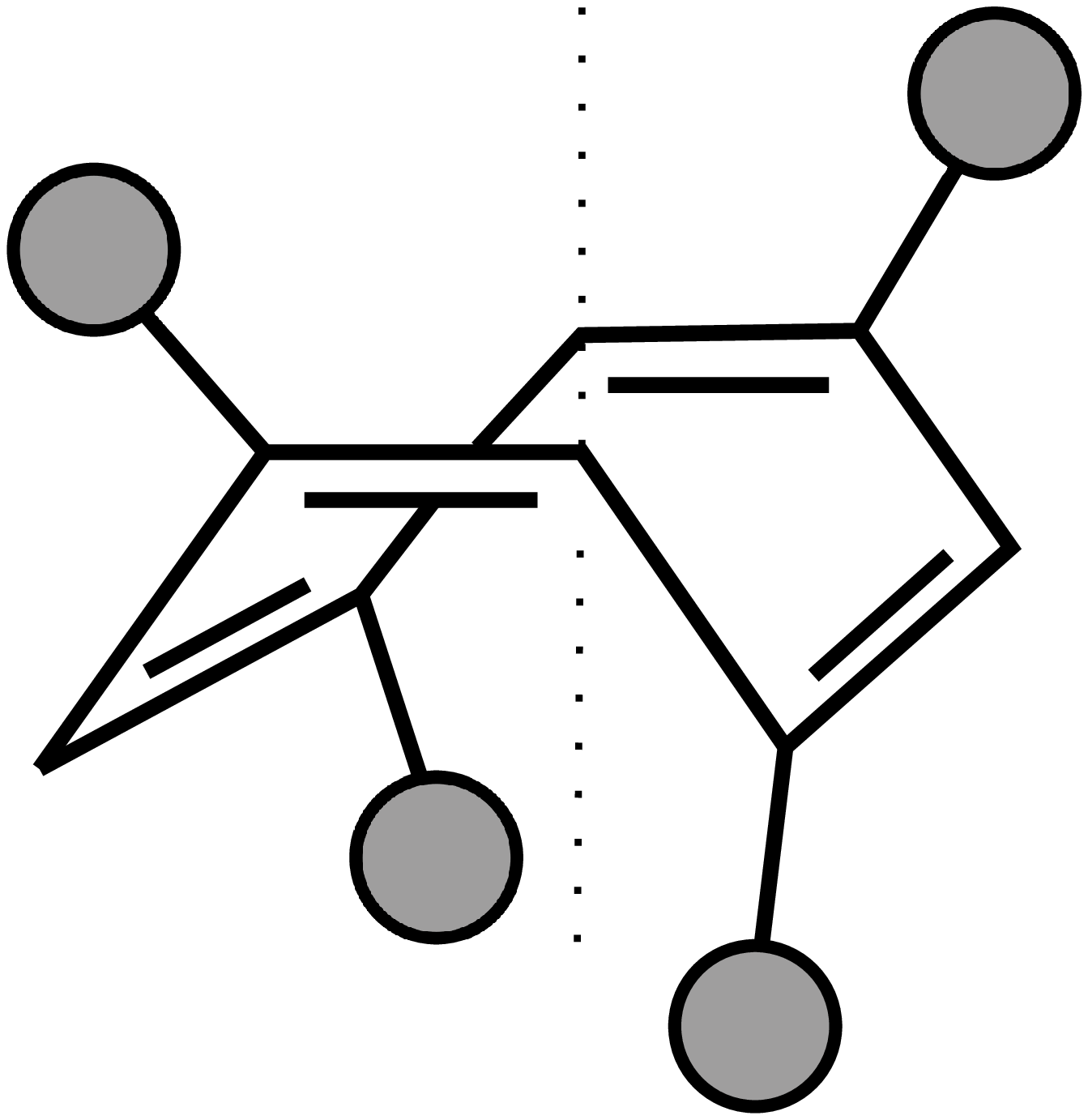}}
\caption{Two examples of achiral molecules.}
\end{figure}
\end{center}

Some particular cases of Eq. \eqref{rotorotation}: if $\beta=0$, then the object has a plane of symmetry \footnote{It should be noted that any 2D planar object is achiral, where the symmetry plane is the plane of the object.}. If $\beta=180$, then the object is symmetric under parity (see Fig. \ref{paritymolecule}). The reason achirality is defined as an invariance relation on $X$ is because the subset of transformations that leave an object $X$ invariant forms a subgroup, the isometry group, and one can categorize the shape of objects (such as molecules) based on the maximal isometry group to which they belong. Such considerations show that for Eq. \eqref{rotorotation}, $\beta=360/n$, where $n$ is a positive integer. \\
 
A useful way to think about all this is to view a reflection acting on a chiral object as creating a new object $\ma R_{\theta,\phi}X$ that lives in a mirror world, where further rotation $R_{\theta',\phi'} \ma R_{\theta,\phi}X$ keeps the object in this mirror world as the determinant is negative one, i.e., $R_{\theta',\phi'} \ma R_{\theta,\phi} \neq R_{\theta'',\phi''} $, and therefore cannot be attained physically with rotation. . 
Upon another reflection one can exit the mirror world, as a product containing two reflections has determinant positive one, i.e.,  $ \ma R_{\theta'',\phi''}R_{\theta',\phi'} \ma R_{\theta,\phi}X=R_{\theta''',\phi'''}X$. A common demonstration is to show that a left-handed glove, when turned inside-out, becomes a right-handed glove: both a reflection in the z-direction (inside-out) and a reflection in the x-direction (left-right) move you into the mirror world, where they are related by rotation. Indeed, left-right, top-down, and front-back reflections are all related by rotations.\\

We note that Eq. \ref{rotorotation}, which takes the form $X= \hat{O}X$, is abstracted to define symmetries in physical law. The set of all transformations $\hat{O}$ that leaves the law of physics $X$ invariant, forms a group. For example, in classical physics, $\hat{O}$ represents the Galilean group which includes translations, rotations, and boosts, and $X$ can represent Newton's laws so that $\hat{O}X$ are Newton's laws seen in the translated, rotated, and boosted frames, respectively. Indeed, modern physics is often done by selecting a set of symmetries $\hat{O}$ that we believe nature respects, and finding laws of physics $X$ that do not change under the symmetry transformations.

\section{Symmetries and Interactions}\label{consequences}

Now that we've defined chiral objects, we move on to interactions. But before we do,  we note that chirality is important in chemistry, where for example the left-handed molecule limonene gives an orange its characteristic smell while its right-handed version gives a lemon its smell \cite{christmas2}. The fact that we can olfactorily distinguish an orange from a lemon means that the detector in our nose, which interacts with the molecule, is itself handed \footnote{Reflecting a handed object along with an achiral detector produces the opposite-handed object with the same detector which, as we will see, if the laws are symmetric under reflection, will produce the same measurement so that one cannot distinguish the chirality of the object.}. We have symmetry of physical law under reflection only if we swap the handedness of both the molecule and detector. Stated another way, only relative chirality can be detected in an interaction. An analogy in physics is the difference in binding between a square knot and a granny knot, which we now proceed to discuss. \\

\begin{center}
\begin{figure}[h]

\subfloat[A square knot, with a right-handed helix on the bottom and a left-handed helix on top. From left to right, the tension jumps from $0$ to $T_L$ as vertex A is crossed, and from $T_R$ to $0$ as vertex B is crossed. \label{reefer}]
{
\includegraphics[scale=0.5]{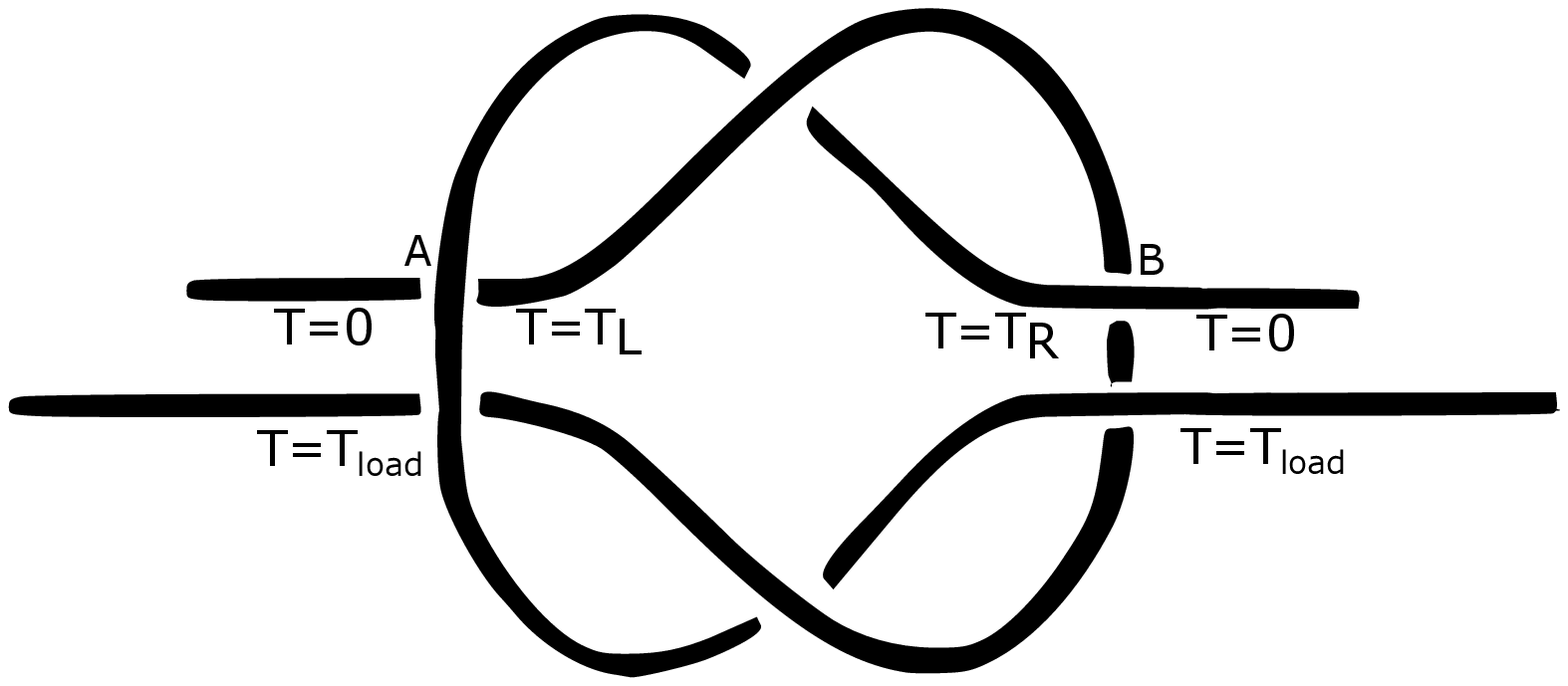}
}
\subfloat[The granny knot, with a right-handed helix on the bottom and a right-handed helix on top. Although the top is the mirror image of the top of Fig. \ref{reefer}, the bottom is not, so the entire knot is not the mirror image of Fig. \ref{reefer}. \label{granny}]
{
\includegraphics[scale=0.5]{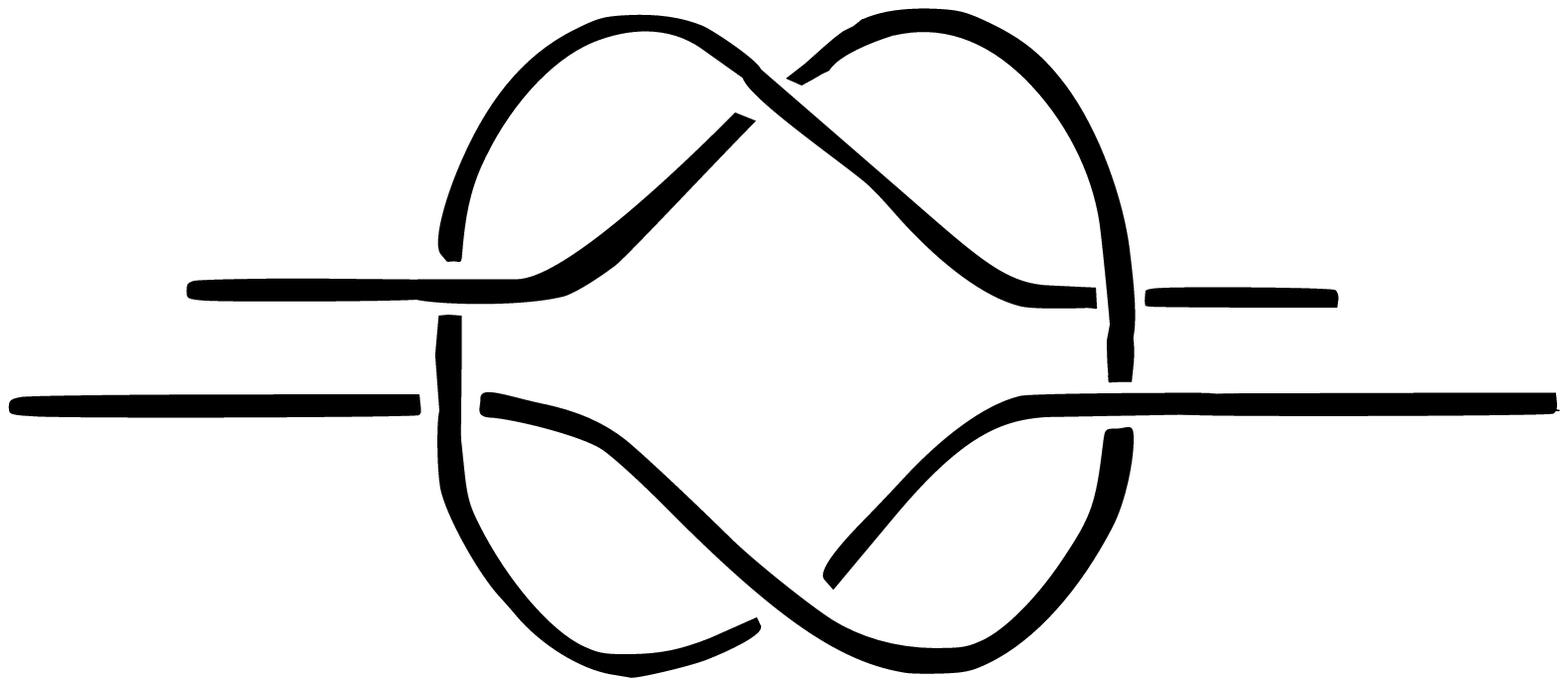}
}

\subfloat[The mirror image of Fig. \ref{reefer}. Due to the invariance of classical physics under reflection, this knot performs the same as Fig. \ref{reefer}. \label{reversereefer}]
{
\includegraphics[scale=0.5]{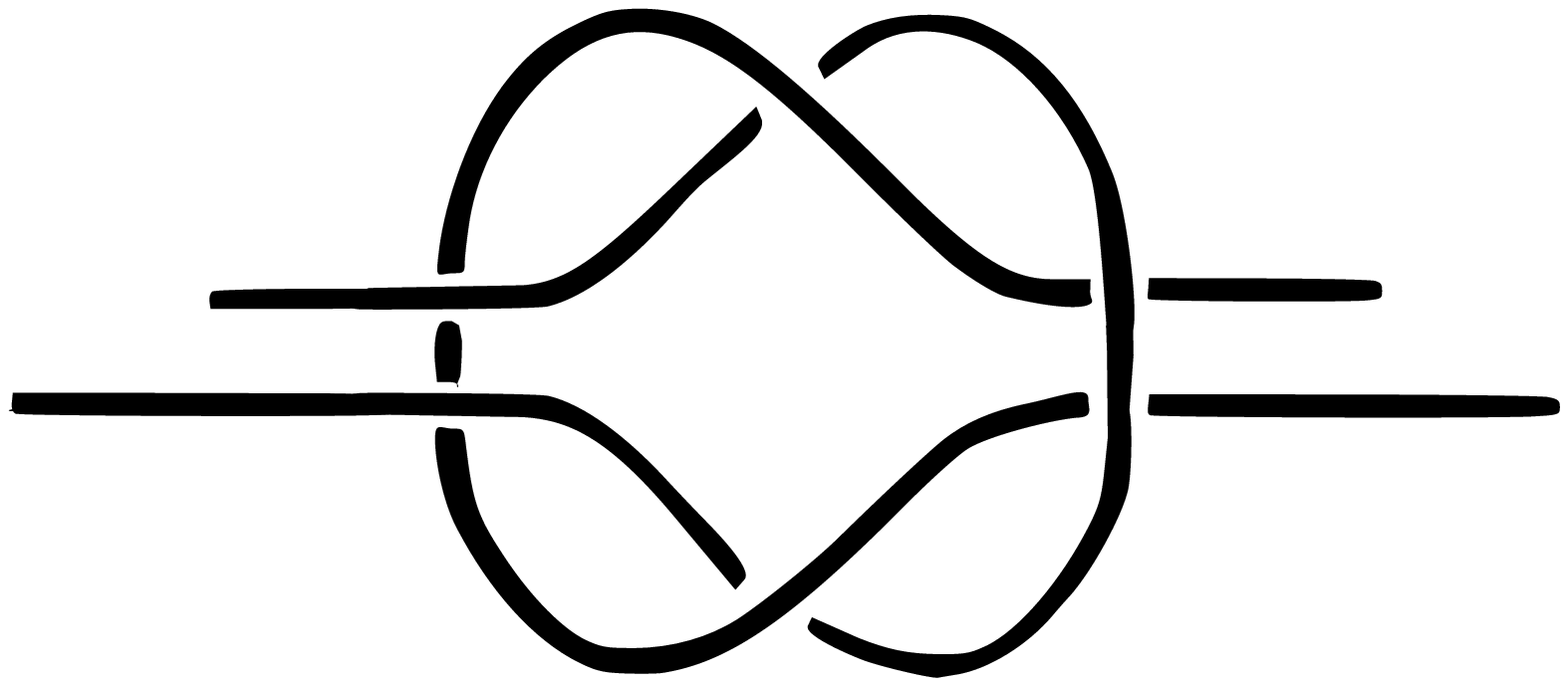}
}

\caption{Although the square and granny knot look similar, they are not mirror reflections of each other, and therefore can (and do) behave differently.}
\end{figure}
\end{center}

The square or reef knot (Fig. \ref{reefer}), which is most commonly used to tie shoe-laces, binds two strings (or two ends of the same string) together by having a twist of one chirality followed by one of opposite chirality \footnote{To tie this knot a common mnemonic is ``right over left, left over right, makes a reef knot both tidy and tight.'' }. By contrast, the granny knot (Fig. \ref{granny}) has two twists of the same chirality \footnote{To tie this knot ``right over left'' is followed by another ``right over left.''}. Both of these knots look mechanically similar, but it is well-known that the granny knot unravels more easily, although only recently was this explained theoretically in detail \cite{twist}. Given that the difference between the two knots are the relative chirality of the top and bottom twists (like-unlike vs like-like), we choose to frame discussion as an interaction between two chiral objects.

We begin by considering the unravelling of the top helix in Fig. \ref{reefer} due to $T_\text{load}$. Starting from vertex A, friction from the left rope (i.e., the rope whose free end is on the left) of tension $T_L$ wound around the right rope at vertex B will allow the tension at the loose end of the right rope (i.e. to the right of B) to be zero without this rope sliding to the left due to the tension $T_R$, thereby unbinding. To calculate the friction needed to prevent slipping we approximate the right rope at vertex B as a pulley around which the left rope is wound (see Fig. \ref{capstanfig}), and use the famous capstan formula  for the friction of a rope wound around a pulley with a wrap angle of $180$ \footnote{For different methods of approximation and analysis, see \cite{bayman,PhysRevLett.115.118302}.}. The friction is $T e^{\pi \mu }-T$, which is derived in the appendix. We get

\begin{center}
\begin{figure}[h] 
\includegraphics[scale=.6]{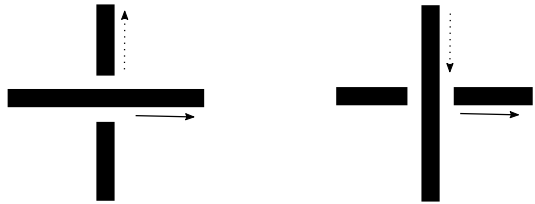}
\caption{A horizontal rope pulled to the right (solid arrow) induces a torque (dotted arrow) on the vertical rope due to friction from rubbing. The direction of the torque depends on whether the horizontal rope crosses over or under the vertical rope. \label{twistEasier}}
\end{figure}
\end{center}

\begin{center}
\begin{figure}[h]

\subfloat[Twisting of a square knot.  \label{torreefer}]
{
\includegraphics[scale=0.5]{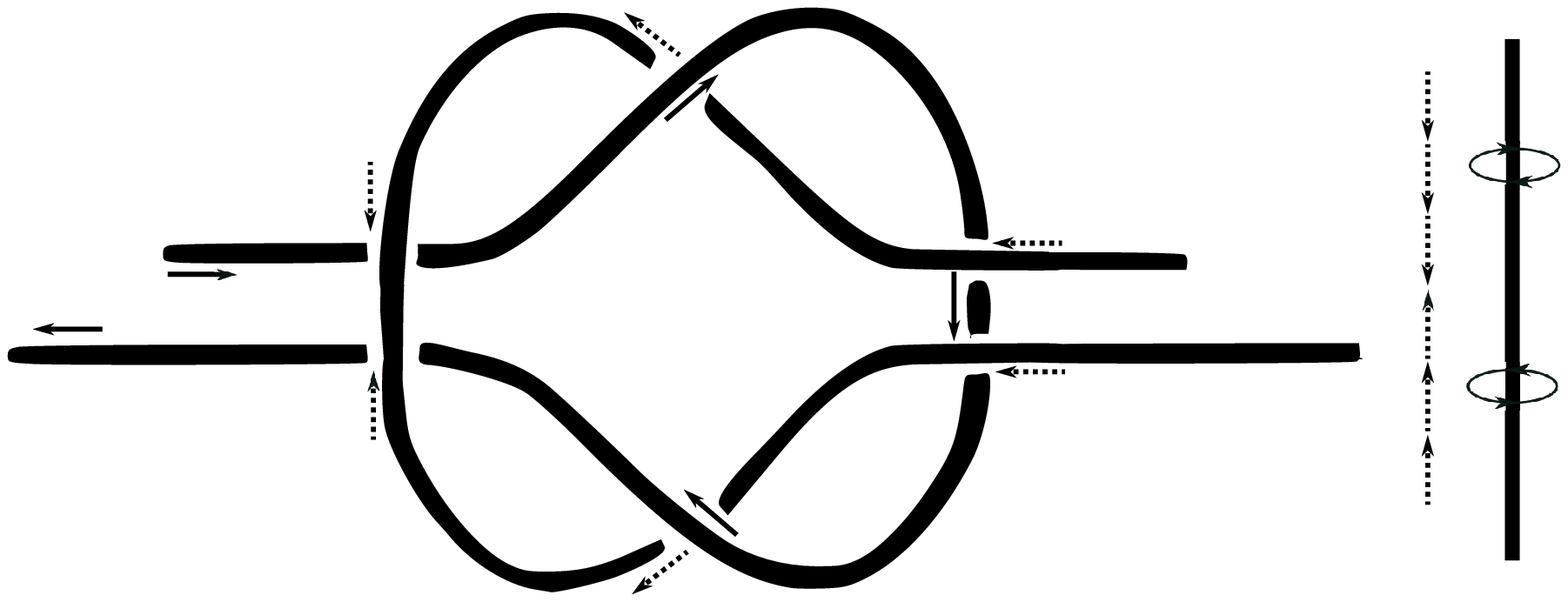}
}
\subfloat[Twisting of a granny knot.  \label{torgranny}]
{
\includegraphics[scale=0.5]{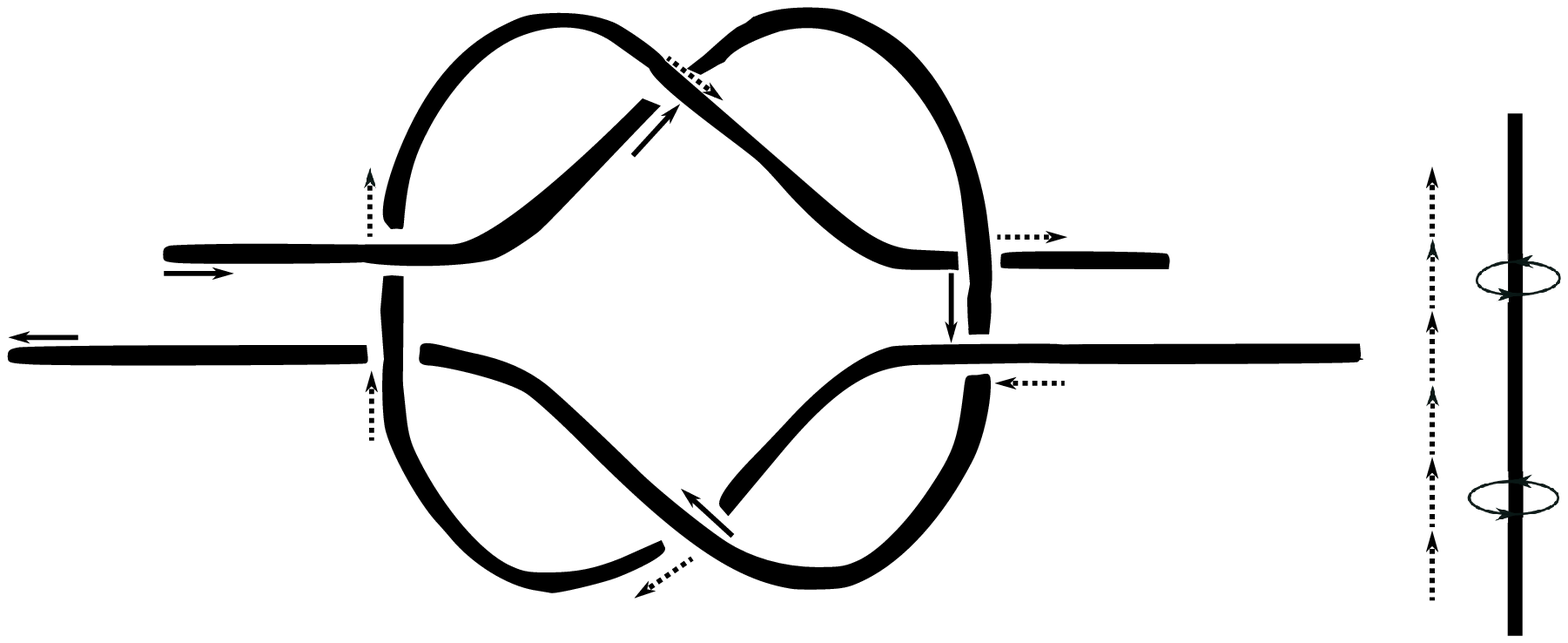}
}
\caption{The solid vector denotes the velocity of parts of the left rope when its bottom end is pulled to the left, while the dotted vector is the local twist on the right rope caused by friction due to the rubbing of the left rope on the right rope due to the initial pulling motion. The image to the right of each knot is the right rope unwrapped to show the induced torsion (or lack thereof).} \label{twister}
\end{figure}
\end{center}

\en
T_R-0&\leq T_L e^{\pi \mu}-T_L\\
\mu&\geq .22,
\een

where we used the symmetry of the square knot to set  $T_L=T_R$. This is in good numerical agreement with \cite{ropes}. The same analysis would give the same value of friction for the granny knot. However, this only considers friction from sliding through the tightened loops and not the twisting of each strand due to friction at the vertices. We will find that such twisting can be seen as an interaction between the top and bottom helices. To illustrate this, consider Fig. \ref{twistEasier}, which shows the torque exerted at a vertex, with the local twisting of the rope given by the dotted arrow using the right-hand rule.  In Fig. \ref{twister}, the bottom of the left rope is pulled to the left and the resulting velocity of the left rope at various points is given by the solid arrow. The resulting torque (dotted arrow) on the right rope  will twist this rope locally in the direction given by the right-hand rule. One can see from Fig. \ref{twister} that for the square knot, the net effect is a clockwise twist on the top part of the rope and a counterclockwise twist on the bottom part of the rope. This type of strain, where two ends are twisted in opposite directions, is known as torsion \footnote{Torsion is similar to tension where a bar is subjected to opposite pulling forces at its ends, except the opposite forces are replaced by opposite torques. A rotational version of Hooke's law is obeyed that tries to restore the system to its original angle, which provides additional stability to the square knot.}, and this resists the initial pull velocity. The granny knot is twisted in the same direction at both ends, so it does not cost anything energetically for the rope to start slipping.

That the knots in Fig. \ref{reefer} and Fig. \ref{reversereefer} have the same strength is due to symmetry of classical physics under reflections. We would say that the same laws of physics are obeyed in the mirror world, by which is meant if we were to actually construct Fig. \ref{reversereefer} in the real world and pull it, then the result would be the same as Fig. \ref{reefer} viewed in a mirror. More generally, if the laws of physics are symmetric under a transformation, then one can perform the transformation first and then time evolve, or time evolve and then perform the transformation: time evolution commutes with the symmetry transformation. This interpretation of symmetry is well-known in both classical and quantum mechanics \cite{shankar}.\\
 
\begin{center}
\begin{figure}[h] 
\includegraphics[scale=0.3]{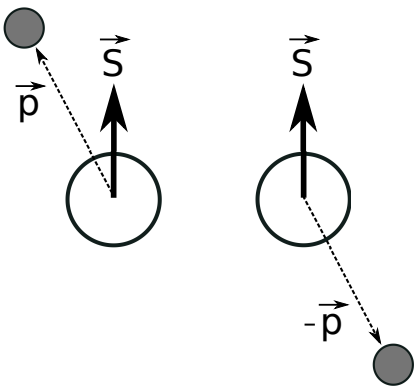}
\caption{Decay of spin-up colbat-60, where the electron is represented by the small circle, and the antineutrino is not shown. The image on the right is related to the one on the left by parity. However, the right image, if realized in the physical world, would have a different probability. Therefore party is violated. \label{wuex}}
\end{figure}
\end{center}

An example of a symmetry not being obeyed is the parity violation experiment of Wu et al \cite{wu}. A spin up $\text{Co}^{60}$ decays via the weak interaction, ejecting an electron: see Fig. \ref{wuex}. Spin is invariant under parity, but momentum reverses sign, so that the mirror image of $\text{Co}^{60}$'s ejection of an electron with momentum $\vec{p}$ is its ejection of an electron with momentum $-\vec{p}$. Therefore a difference in the decay rate between $\vec{p}$ and $-\vec{p}$ in the real world would indicate a violation of parity \footnote{See \cite{weak} for a detailed calculation of the $\text{Co}^{60}$ experiment.}, and such a difference is indeed found. This stunning result, predicted by Lee and Yang \cite{leeyang}, shows that as far as weak interactions are concerned, symmetry arguments such as those used to equate the knots of Fig. \ref{reefer} and Fig. \ref{reversereefer} are invalid and instead each knot must be checked individually, as was done above.

It was mentioned that chirality can only be detected by a chiral detector \footnote{An example of a chiral detector is an enzyme whose shape precludes it from binding to a molecule of the opposite-handedness.}, and if one were to swap the chirality of both the object and detector, the same result would occur: in other words, only relative chirality can be detected. However, the violation of parity means one can detect chirality absolutely: left and right can be distinguished in an absolute sense. 

\section{Conclusions}\label{conclusions}

The importance of chirality is reflected in the fact that it receives its own section in introductory organic chemistry \cite{wade}. By contrast, in physics, examples of chiral objects are scattered. It is only in quantum mechanics where some formal organization is given to the topic with the introduction of the parity operator. In this paper we argued the merits of having a dedicated section on this topic much earlier than quantum mechanics, to provide a connection between otherwise disparate topics such as the inversion of a plane mirror, solenoid handedness, various right-hand rules, circular polarizations, and threaded ropes and fasteners. We have also shown that reflections are a particularly simple transformation, which makes them ideal for serving as preparation for more abstract concepts such as similarity transformations, group actions, and symmetry in physical law with its connection to commutativity. Both mirrors and knots have the additional merit that they visually display chirality. Finally, the fact that not just certain objects, but that nature itself is chiral (due to the weak interaction), has intellectual interest beyond particle physics.


\section{Appendix: The Capstan Equation}

\begin{center}
\begin{figure}[h] 
\includegraphics[scale=1]{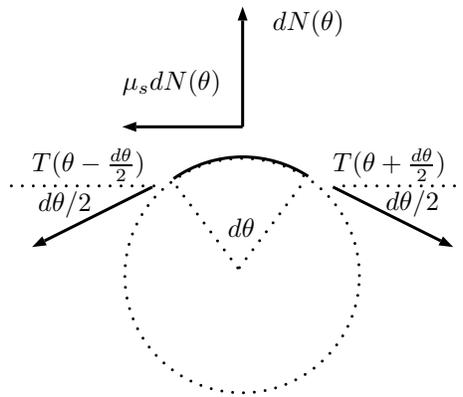}
\caption{A rope wrapped around a pulley. The rope in the figure represents the left rope in Fig. \ref{reefer}, and the pulley in the figure represents the circular cross-section of the right rope in Fig. \ref{reefer} at vertex B. The size $d\theta$ has been exaggerated for clarity.  \label{capstanfig}}
\end{figure}
\end{center}

To find the tension in a cord wrapped around a rough pulley, equilibrium requires that the sum of the forces in the horizontal and vertical directions is zero (see Fig. \ref{capstanfig}). Taylor expanding the tension and keeping terms only to order $d\theta$ one gets:

\en \label{eqn11}
\f{dT}{d\theta}d\theta-\mu_s dN&=0\\
dN-T d\theta&=0.
\een

Solving the second line for $dN$ and substituting into the first line gives

\en \label{eqn12}
\f{dT}{d\theta}&=\mu_s T\\
T(\theta)&=T(0)e^{\mu_s \theta},
\een

which is the famous capstan equation \cite{capstan,cap2}, which states that a tension of $T(0)e^{\mu \theta}$ at one end can be supported with a smaller tension $T(0)$ at the other end, where $\theta$ is the wrap angle around the pulley.\\

From Eqn. \eqref{eqn11} and Eqn. \eqref{eqn12}, we now have the distribution of normal force $dN(\theta)=T(0)e^{\mu_s \theta}d\theta$ of the left rope wound around the right rope, so that a maximum bound can be placed on the friction exerted on the right rope:


\en
f&=\int   \mu_s\, dN(\theta)\\
&=T(0)e^{\pi \mu}-T(0).
\een

We emphasize that this frictional force is what prevents the pulley in Fig. \ref{capstanfig} from sliding into or out of the page: the tighter the rope is wound around the pulley, the more difficult it is for the pulley (which represents the cross-section of the other rope) to slide beneath the rope.
\end{document}